\documentclass[11pt,english,pre,showpacs,amsmath,amssymb,preprint]{revtex4}
\usepackage[T1]{fontenc}
\usepackage[latin1]{inputenc}

\makeatletter

\usepackage{graphicx} 
\usepackage{bm} 

\newcommand{\be}{\begin{eqnarray}}
\newcommand{\ee}{\end{eqnarray}}

\usepackage{babel}
\makeatother
\begin{document}

\title{Growth of Order in An Anisotropic Swift-Hohenberg Model}

\author{Hai Qian and Gene F. Mazenko}

\affiliation{The James Franck Institute and Department of Physics, The University
of Chicago, Chicago, Illinois 60637}

\date{Sep. 10, 2005}

\begin{abstract}
We have studied the ordering kinetics of a two-dimensional anisotropic
Swift-Hohenberg (SH) model numerically. The defect structure for this
model is simpler than for the isotropic SH model. One finds only dislocations
in the aligned ordering striped system. The motion of these point
defects is strongly influenced by the anisotropic nature of the system.
We developed accurate numerical methods for following the trajectories
of dislocations. This allows us to carry out a detailed statistical
analysis of the dynamics of the dislocations. The average speeds for
the motion of the dislocations in the two orthogonal directions obey
power laws in time with different amplitudes but the same exponents.
The position and velocity distribution functions are only weakly anisotropic.
\end{abstract}


\maketitle

\section{Introduction}

There is on going interest in the growth kinetics of stripe forming
systems. There has been progress via experimental \cite{Harrison,Harrison2}
and numerical \cite{CB98,HQ1} studies of growth after a quench from
an isotropic initial state. However the theoretical understanding
of such systems remains limited. This is mostly due to the complexity
of the defect structures generated during ordering in such systems.
For example, in the Swift-Hohenberg model, there are grain boundaries,
disclinations and dislocations generated in the ordering process.
The co-existence of all these different defect structures has hindered
the theoretical analysis of the striped phase ordering systems. In
this paper, we study an anisotropic Swift-Hohenberg (SH) model, where
only dislocations are produced in the ordering process. Our goal is
to understand the statistical properties of these defects much as
we now understand those properties for simple vortex producing models.

There are formal arguments\cite{PismenASH} that if we break the symmetry
of the isotropic SH model by applying, for example an electric field,
then the system can be mapped onto an an isotropic TDGL model. This
suggests a $L\approx t^{1/2}$ growth law compared to much slower
growth in the isotropic SH model. We find support for this hypothesis.

Some previous studies have focused on the evolution of a few dislocations
\cite{Cross,Goren,Braun,struc,BP,kramer,rasenat}. Tesauro and Cross
\cite{Cross} studied the steady state climbing motion (move along
the direction of stripes) of isolated dislocations both theoretically
and numerically in several two-dimensional model systems including
the SH model. They found that the wave number selected by dislocation
climb is marginally stable only for potential models. Bodenschatz
et al. \cite{struc} studied the climbing motion of dislocations with
amplitude equations appropriate for systems with an axial anisotropy.
The Peach-Kohler (PK) force ( the effective wave number mismatch)
drives the dislocation motion, just as in Ref. \cite{Cross}. They
also consider the interaction between two dislocations together with
the PK force. Goren et al. \cite{Goren,rasenat,kramer} studied the
convection in a thin layer of a nematic material experimentally. They
introduced a gauge-field theoretical treatment to study the climbing
of dislocations in a stressed background field where the PK force
plays a role. The theory \cite{BP} predicts that climbing and gliding
motions of a single dislocation are equivalent (after the proper scaling
for the anisotropic system) and due to the PK mechanism. Braun and
Steinberg \cite{Braun} studied the same experimental system. They
measured the gliding motion of dislocations due to a pure interaction
between the members of the pair without the PK mechanism. They found
that the climb and gliding motion have different characters.

Boyer \cite{boyer} simulated an anisotropic stripe forming model
\cite{pesch} based on the Swift-Hohenberg model. His model is more
complicated than ours. In his model the stripes have two preferred
directions and a zig-zag pattern is formed, and the dislocations tend
to stay together to form large domain walls. The author found that
for small quenches the energy, the dislocation energy and the characteristic
length normal to the stripes all scale as $t^{\pm1/2}$ ($+$ for
the characteristic length). He also found that for deep quenches the
system was frozen. The pinning effect becomes important as the quench
depth increases. The zig-zag pattern was experimentally realized in
Ref. \cite{kamaga}.

Here we study an ensemble of well separated dislocations in the context
of domain growth. The motion of the dislocations in this model is
highly anisotropic. They tend to move across the stripes. The average
speeds across and along the stripes obey simple power laws in time
with different amplitudes but approximately the same exponent. The
distributions of the defect velocities along the two orthogonal directions
have same form and large velocity power-law tails with approximately
the same exponents. Two bulk measurements of the ordering, the decay
of the effective energy and the number of dislocations, obey a simple
power law in time with a logarithmic correction, as for the XY-model
\cite{HQ2}.

The two dimensional isotropic Swift-Hohenberg (SH) model \cite{SH77}
is defined by a Langevin equation 
\begin{equation}
\frac{\partial\psi(\mathbf{x},t)}{\partial t}=-\frac{\delta\mathcal{H}[\psi]}{\delta\psi(\mathbf{x},t)}+\xi(\mathbf{x},t)\,\,,\label{SH}
\end{equation}
where $\psi$ is the ordering field, and the effective Hamiltonian
is given by 
\begin{equation}
\mathcal{H}[\psi]=\int d^{2}r\,\left\{ -\frac{\epsilon}{2}\psi^{2}+\frac{1}{2}\left[(\nabla^{2}+1)\psi\right]^{2}+\frac{1}{4}\psi^{4}\right\} \,\,,\label{H}
\end{equation}
 where $\epsilon$ is a positive constant. All the quantities in this
paper have been put in dimensionless form. The noise $\xi$ satisfies
$\langle\xi(\mathbf{x},t)\xi(\mathbf{x}',t')\rangle=2T\delta(\mathbf{x}-\mathbf{x}')\delta(t-t')$,
where $T$ is the temperature after the quench. In the following,
we set $T=0$ which eliminates the noise term from the analysis. Starting
from a random initial condition without long distance correlations,
the SH equation (\ref{SH}) generates stripes with period $2\pi$.

In the simulations for the isotropic SH model, we found \cite{HQ1}
that the grain boundaries' motion dominate the ordering dynamics of
the system, which is different from what is seen in some experiments
\cite{Harrison,Harrison2}, where the disclination quadrapole annihilation
is the dominant ordering process. Disclinations and dislocations are
also present in the SH ordering system. In more recent experiments
\cite{Ward} on different diblock copolymer systems, defect configurations
looking more like the SH simulations \cite{HQ1} are found. The co-existence
of different kinds of disordering defects makes it difficult to analyze
simulations of the isotropic SH system. However, in an anisotropic
SH system, where only point-like dislocations are present, the system
should be easier to study.

We make the SH system anisotropic by adding an additional term to
the effective Hamiltonian 
\begin{equation}
\mathcal{H}[\psi]\rightarrow\mathcal{H}[\psi]+\int d^{2}r\,\frac{\gamma}{2}\left(\frac{\partial\psi}{\partial y}\right)^{2}\,\,,\label{gterm}
\end{equation}
 where $\gamma$ is a constant. The anisotropic term corresponds to
applying an external magnetic field. In this case the SH equation
now takes the form 
\begin{equation}
\frac{\partial\psi}{\partial t}=\epsilon\psi+(\nabla^{2}+1)^{2}\psi-\psi^{3}+\gamma\,\frac{\partial^{2}\psi}{\partial y^{2}}\,\,.\label{aSH}
\end{equation}
 We studied the case where $\epsilon=0.1$ and $\gamma=0.4$. Stripes
generated by the above equation align along the $y$ direction on
the $x$-$y$ plane. This configuration minimises the anisotropic
term in Eq. (\ref{gterm}). We start from a random initial condition
for $\psi$. After a very short transient time, the only defects left
in the system are dislocations. Dislocation annihilation is the final
ordering process.

In Sec. II, we set up our numerical study. Then in Sec. III we study
the time decay of the system energy. The stripe patterns and the motions
of the dislocations are shown in Sec. IV. And we analyse the quantitative
measurements in Sec. V. The speed distribution for the dislocations
are shown in Sec. VI.

\section{Numerical Algorithm}

We employed the usual Euler method to drive the system: 
\begin{equation}
\psi(t+\Delta t)=\psi(t)+\Delta t\left(\epsilon\psi(t)-(1+\nabla^{2})^{2}\psi(t)+\gamma\partial_{y}^{2}\psi(t)-\psi(t)^{3}\right)
\end{equation}
 We take in this case time step $\Delta t=0.02$ and lattice spacing
$\Delta r=\pi/4$. In the following sections, the numerical measurements
are obtained from the system that is evolved by the Euler method.

\section{Time Decay of the Average Energy}

The first quantity we look at is a gross statistical measure of the
ordering given by the average coarse-grained energy $E=\langle{\mathcal{H}}\rangle_{t}$.
In Fig. 1 we plot $\Delta E=E-E_{0}$, where $E_{0}$ is the ordered
value of $E$ (known to be accurately given by $E_{0}=-\epsilon^{2}S/6$,
where $S$ is the surface area of the system). These simulations were
averaged over $528$ runs. In agreement with the $n=d=2$ TDGL model
\cite{HQ2,HQ5} we find a power law with a logarithmic correction
characteristic of the annihilation of point defects. This is clearly
consistent with a growth law exponent of $z\approx2$. Although Fig.
1 does not give a good estimate for the exponent $z$, we conclude
that $z=2$ is the best value by taking into account Fig. 6 discussed
below. Having established that the ordering is speeded up relative
to the isotropic SH model, where $z\approx3$, we can move on to look
at the nature of the ordering patterns grown using this model.

\begin{figure}
\begin{center}\includegraphics[scale=.31]{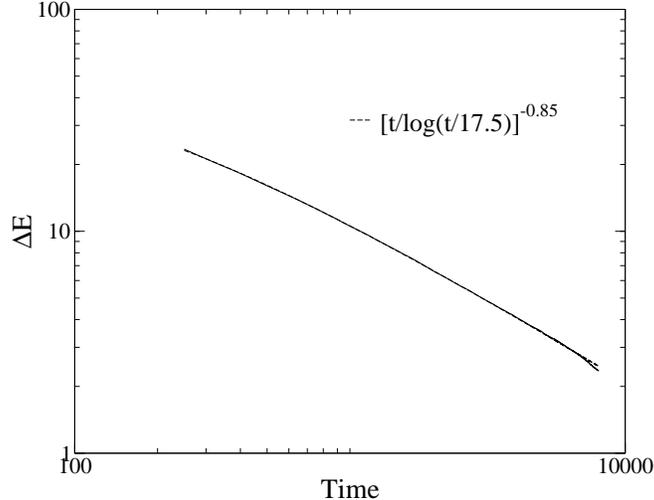}\end{center}

\caption{$\Delta E$ v.s. time $t$ after the quench. The energy $\Delta E$
is proportional to $\left[t/\log(t/18.5)\right]^{-0.85}$. All averages
are of 528 runs.}
\end{figure}

\section{Dislocations}

In Fig. 2 we show a typical ordering configurations of the anisotropic
SH system at different times. Notice that the only defects produced
are dislocations.

\begin{figure}
\begin{center} \includegraphics[scale=.47]{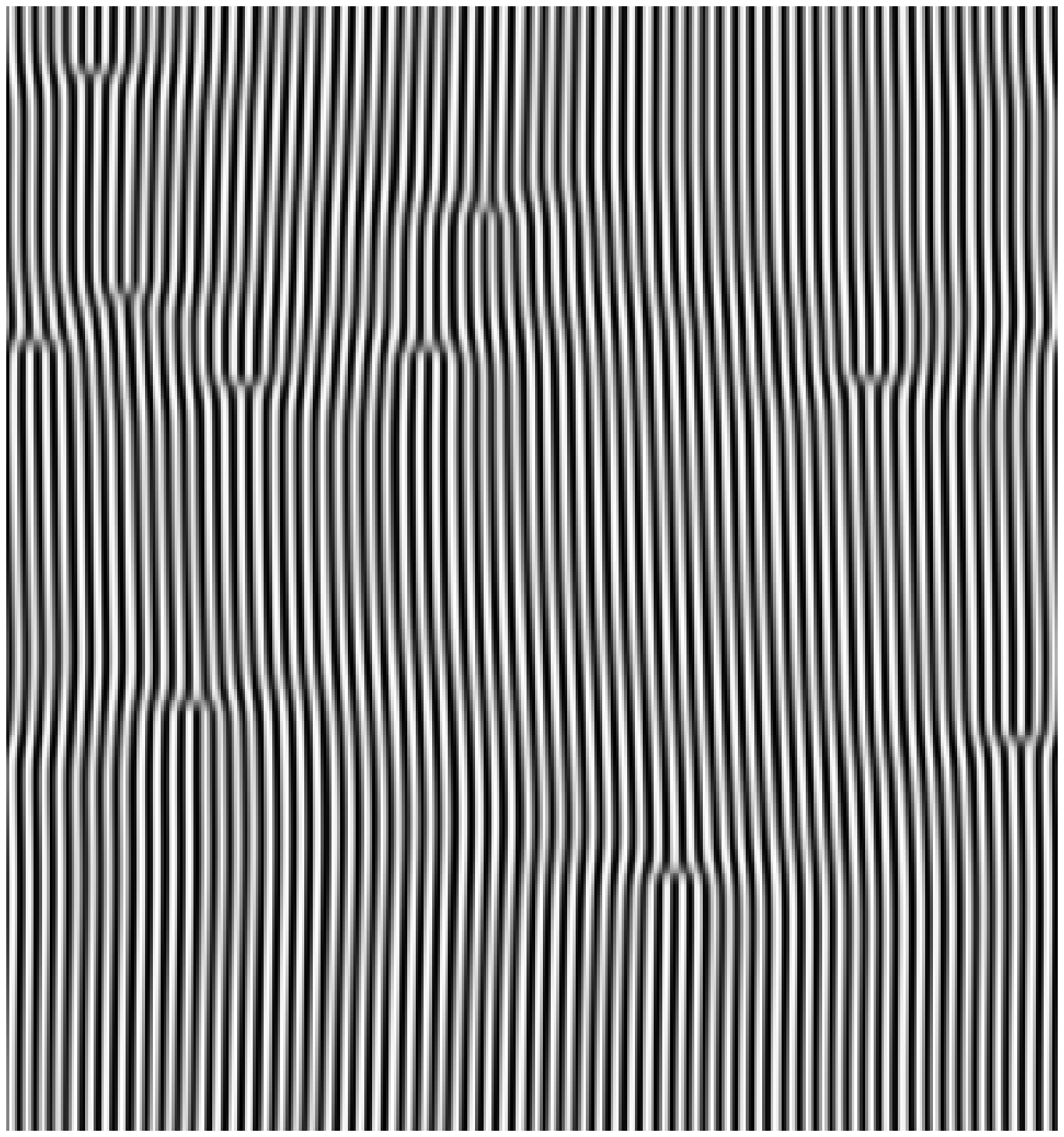} \includegraphics[scale=.47]{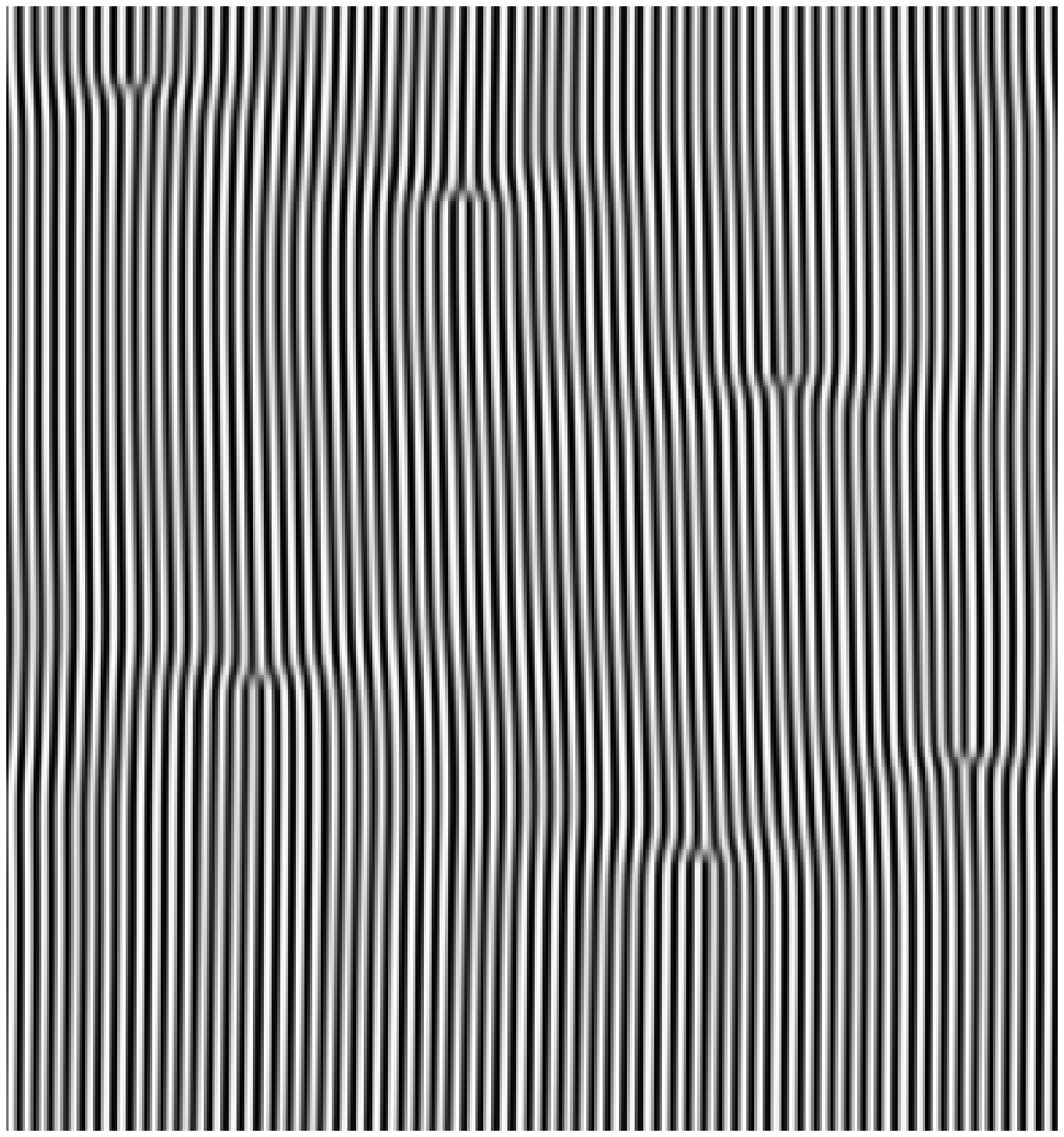} \includegraphics[scale=.47]{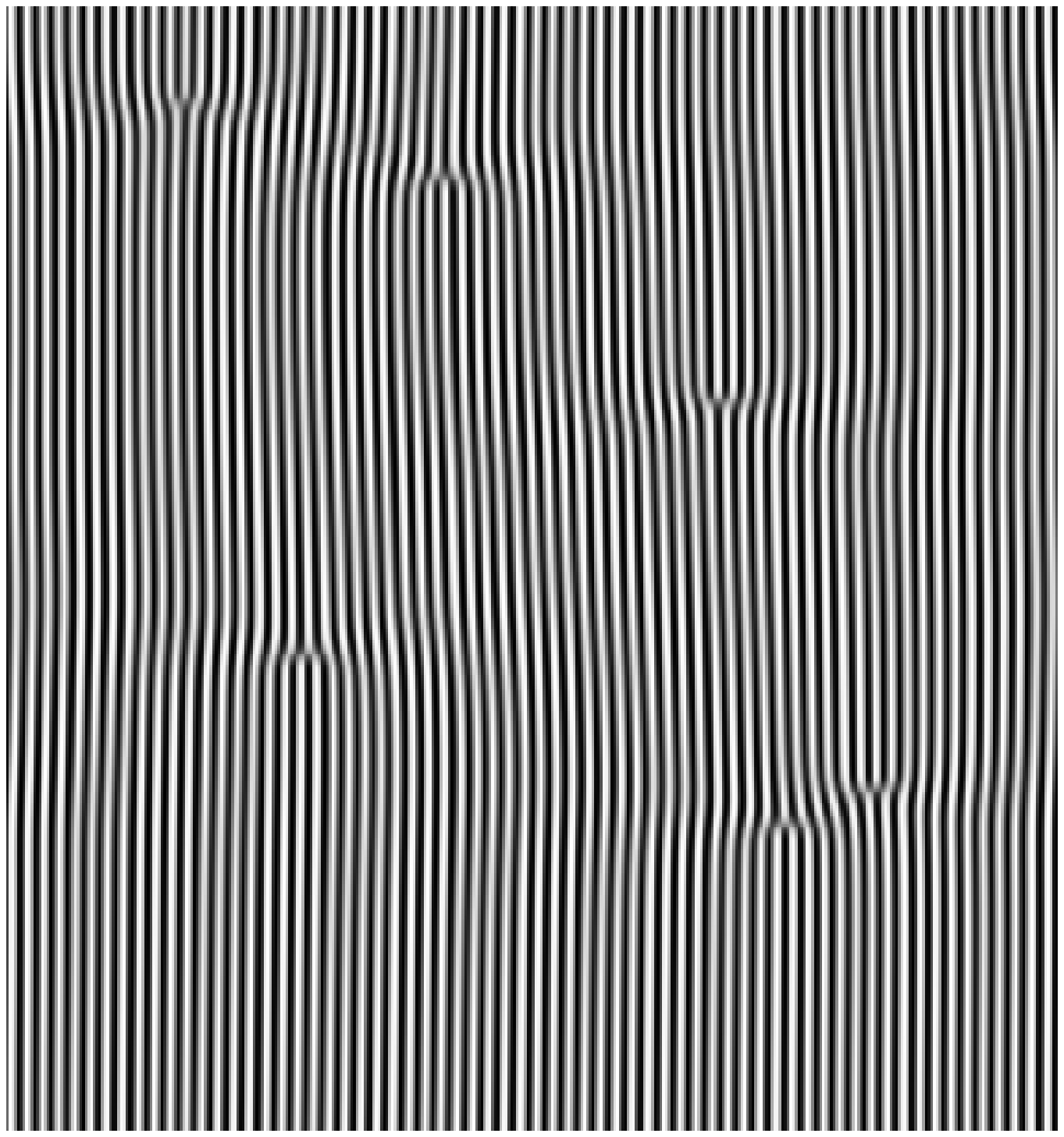} \includegraphics[scale=.47]{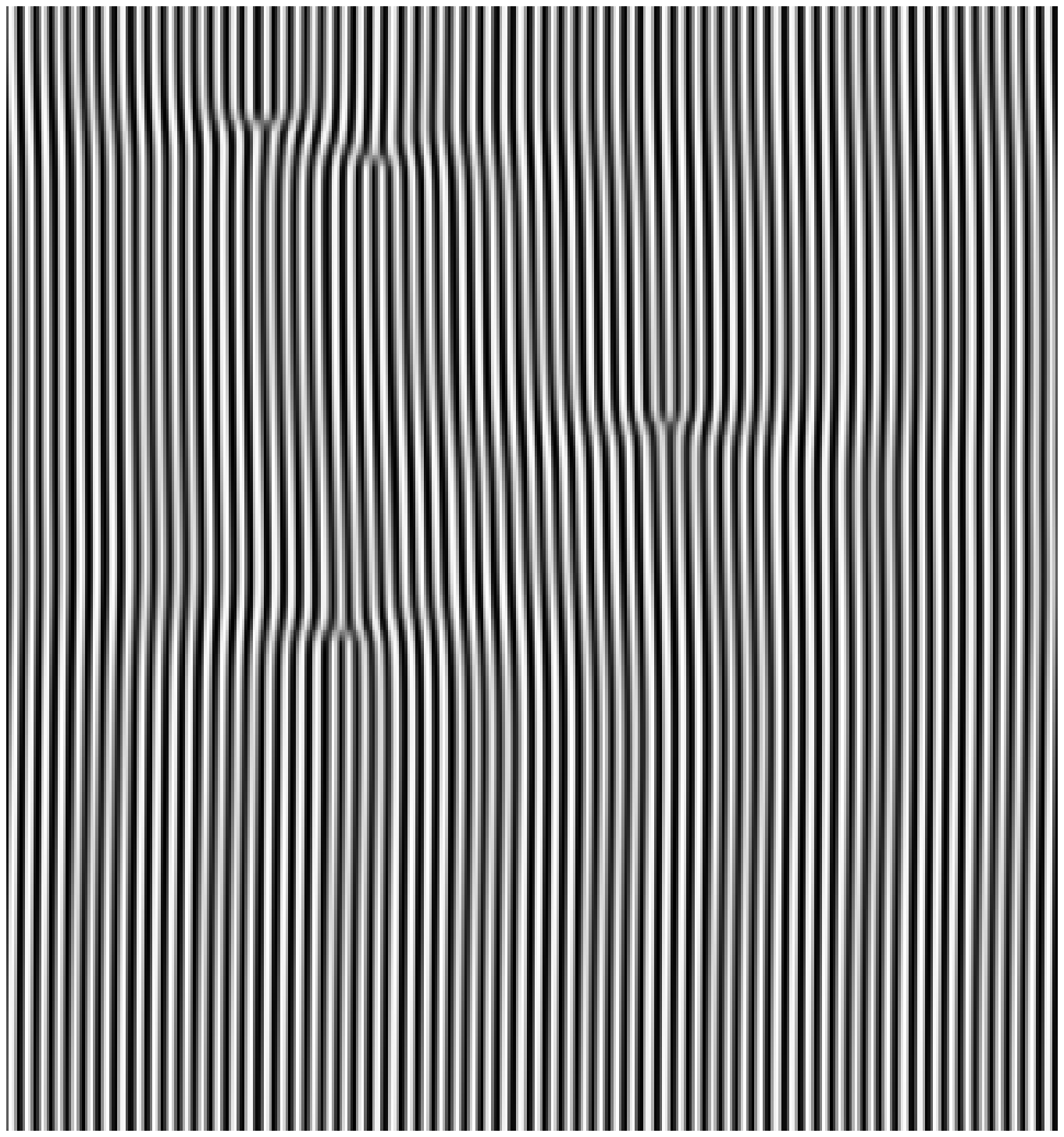} \includegraphics[scale=.47]{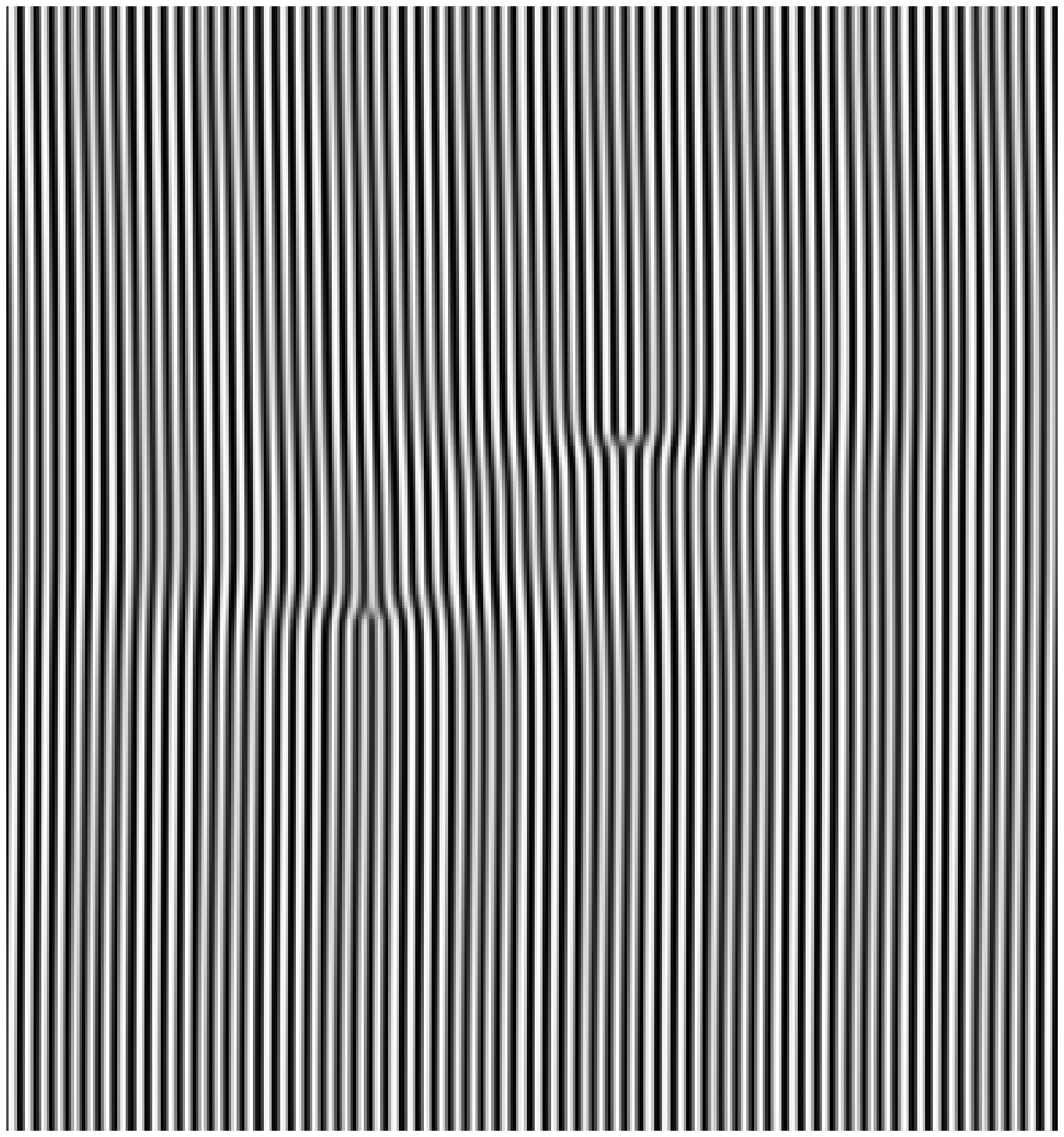} \includegraphics[scale=.47]{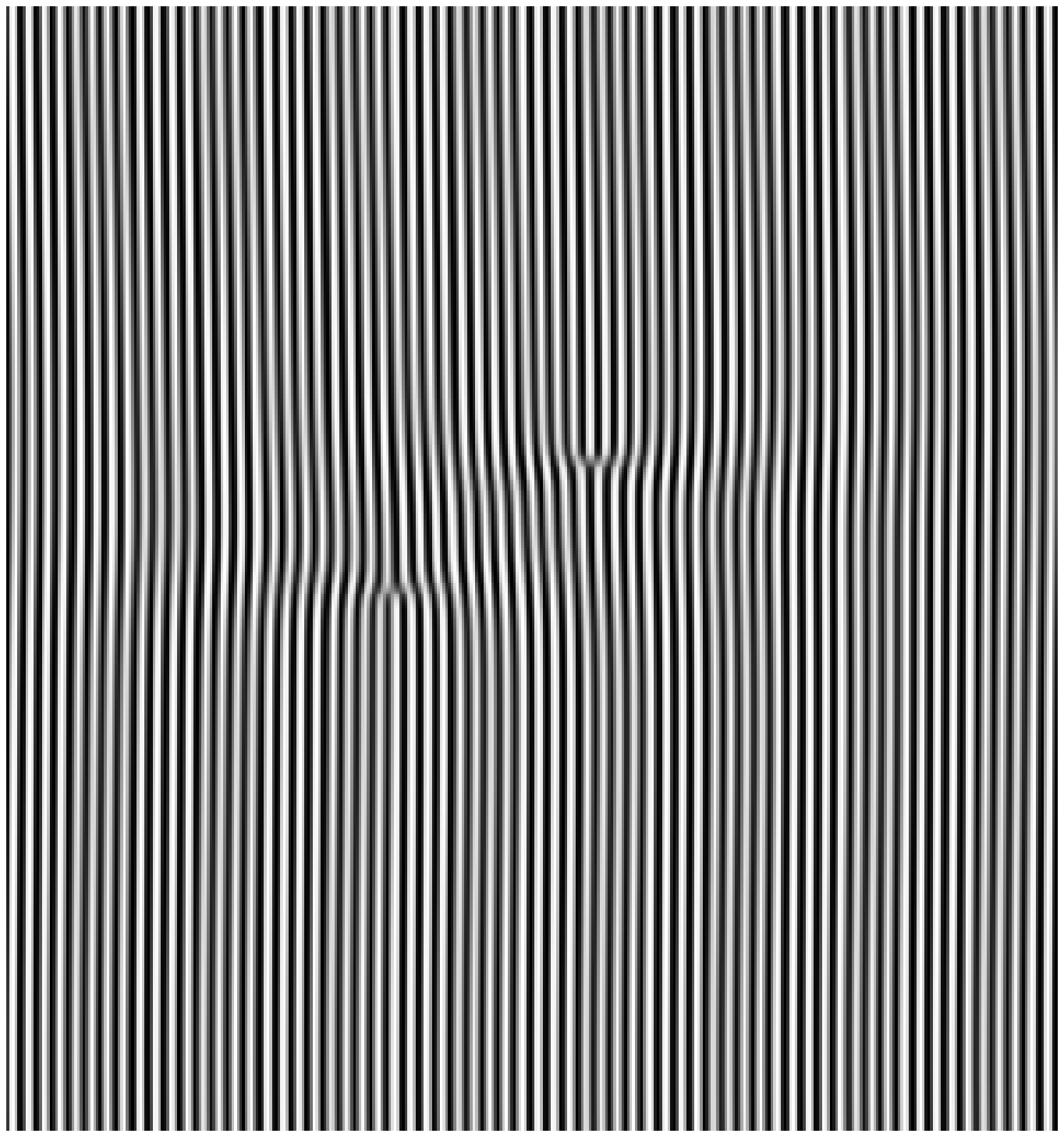}\end{center}

\caption{Typical configurations of an $512\times512$ anisotropic SH system
at different times. From left to right and top to bottom, the system
was at $t=2500$, $5000$, $7500$, $10000$, $12500$ and $15000$.}
\end{figure}

We are interested in the path of each dislocation trajectory. These
paths must be determined accurately enough such that we can compute
dislocation speeds. In the case of the $\mathcal{O}(2)$ TDGL model
\cite{HQ2,HQ5} we were able to accurately determine the position
of a vortex by finding the zeros (minima) in the order parameter amplitude.
Here the situation is more complicated. As explained in reference
\cite{HQ1} the positions of defects in the SH model are located by
maxima in the quantity 
\begin{equation}
A=\sum_{\alpha}\left(\nabla_{\alpha}\phi\right)^{2}
\end{equation}
 where $\phi$ is the angle that the director $\hat{n}=\nabla\psi/\left|\nabla\psi\right|$
makes with some arbitrary direction and $\alpha$ is the index for
different spatial directions. We showed numerically that if $A>A_{0}$
then that site on the lattice can be associated with a defect. Here
we need to determine the position of the dislocation with some accuracy.
We have found that the expression 
\begin{equation}
\bar{r}_{\alpha}=\frac{\sum_{i}r_{\alpha}^{i}A_{i}}{\sum_{i}A_{i}}
\end{equation}
 gives the position of the dislocation in the $\alpha$th direction,
and the sum is over all contiguous sites where $A_{i}>A_{0}=3.0$.
Using these procedures we obtain, for example, the set of dislocation
trajectories shown in Fig. 3. The dislocations tend to move (glide)
across the stripes and annihilate with each other. If we look more
closely we see the oscillating behavior in the glide motion as shown
in Fig. 4 and 5. In Fig. 4 we also show all the sites that are associated
with the dislocations. To obtain useful statistical data we will need
to average over the different runs for our system.

The oscillation in the glide motion of the dislocation is due to Peierls-like
pinning forces. The law of motion of a dislocation due to the Peierls-like
force takes the form \cite{boyer}
\[
\mu^{-1}v=\mu^{-1}\frac{dx}{dt}=f-p\cos(kx)\ .
\]

Here $v$ is the velocity across the stripes. $\mu$ is
the mobility. $f$ is the external force per unit length applied onto
the dislocation. In this context, $f$ is caused by other dislocations
especially the one that is going to annihilate with the dislocation
of interest. $p$ is the magnitude of the pinning force. The Peierls-like
pinning force term oscillates with a period of $2\pi/k$, which is
exactly the stripe pattern period. We are interested in the interaction
$f$ between dislocations. In the next section, we will compute the
the average $v$ versus the separation distance between two annihilating
dislocations.

\begin{figure}
\begin{center} \includegraphics[scale=0.45]{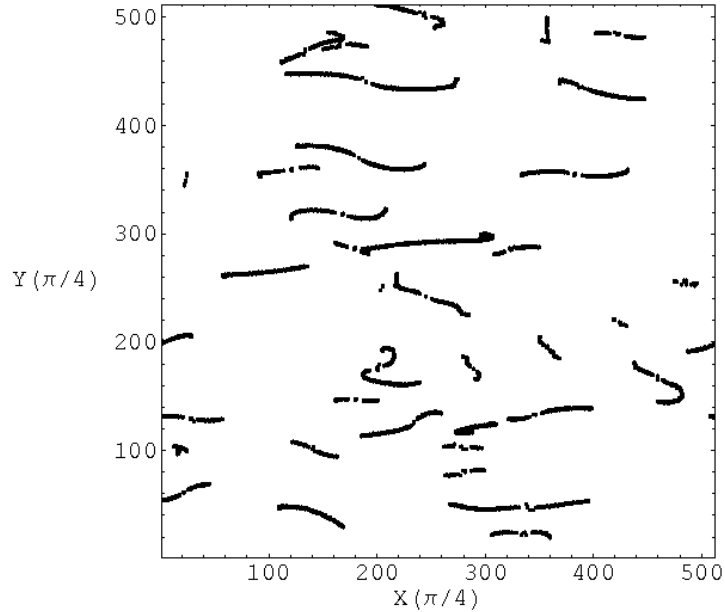}\end{center}

\caption{The trajectories of the dislocations in a $512\times512$ anisotropic
SH system. The dots are the positions where two dislocations annihilate.
The stripes are along the $y$-direction as shown in Fig. 2. Most
of the dislocations glide across the stripes.}
\end{figure}

\begin{figure}
\begin{center} \includegraphics[scale=0.24]{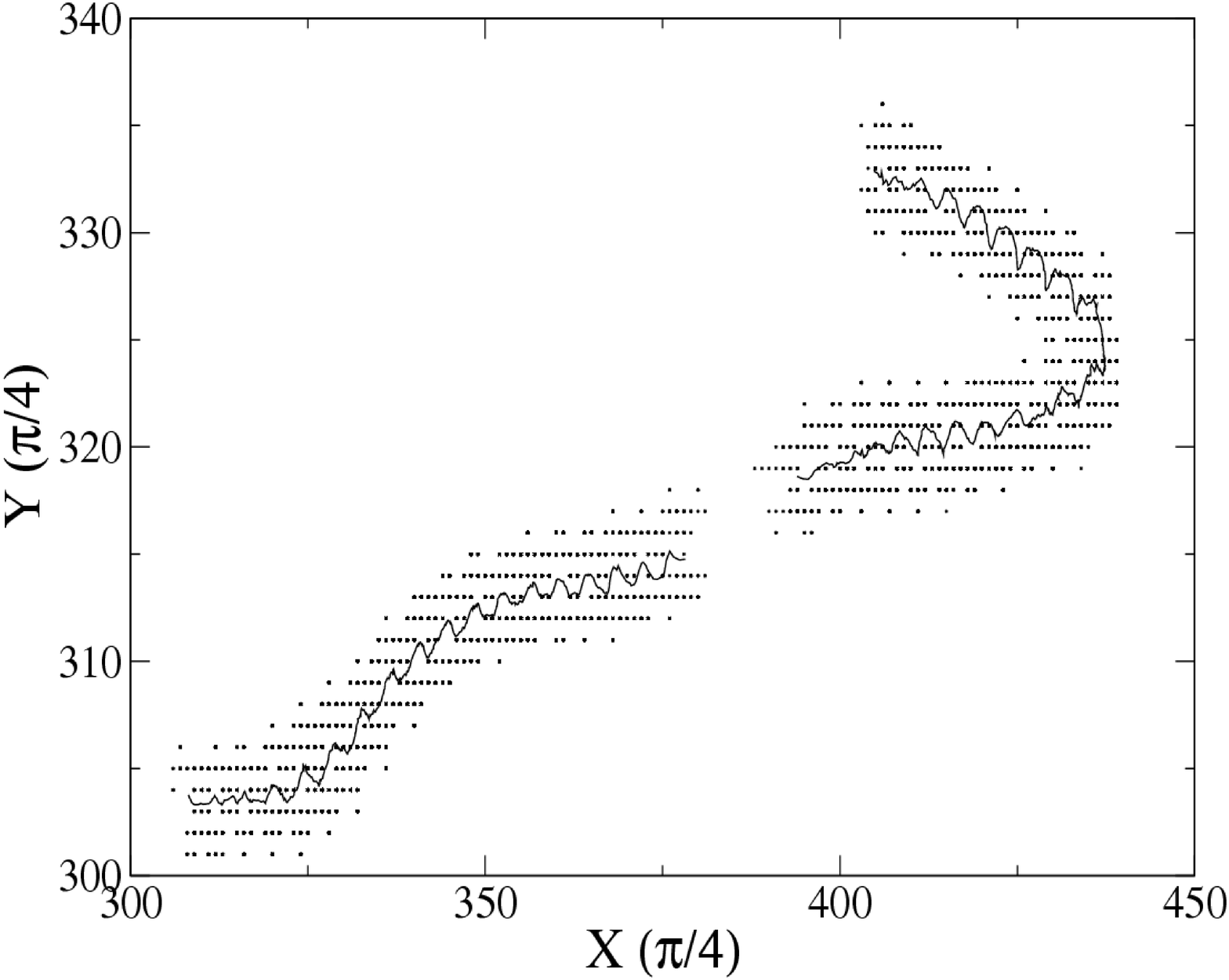}\end{center}

\caption{The trajectories of two annihilating dislocations. The solid line
denotes the center of the dislocation core determined by Eq. (7).
The dots indicate the sites $i$ which are part of the core used in
Eq. (7). Thus this figure shows the motion of the dislocation core. }
\end{figure}

\begin{figure}
\begin{center}\includegraphics[scale=0.39]{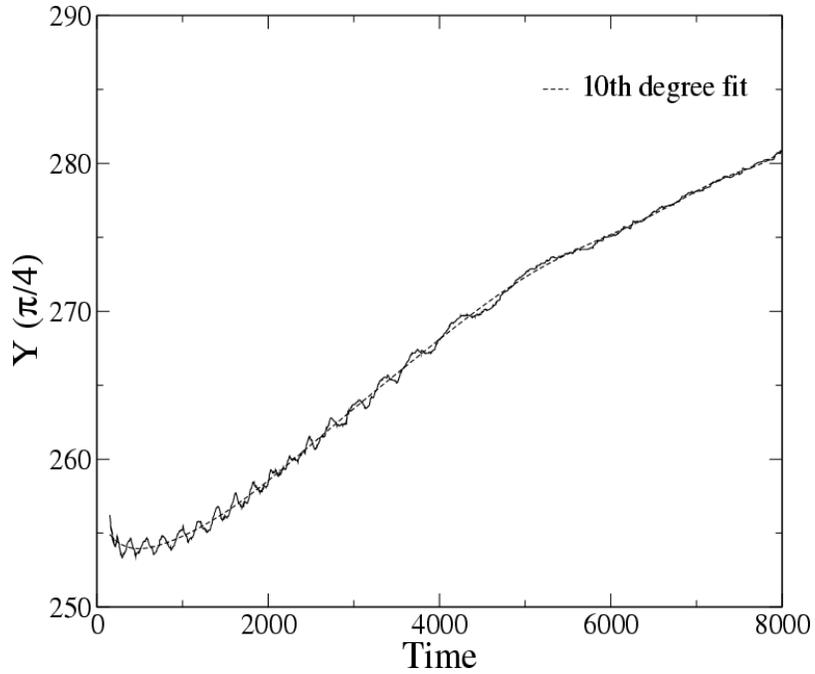}\end{center}

\caption{The position of a dislocation along the stripes versus the time after
the quench. We use a 10th degree polynomial to fit the data and average
out the oscillations along $y$-direction.}
\end{figure}

We plot the average number of dislocations $N_{d}$ as a function
of time in Fig. 6. Clearly it is fit by a power law with a log correction
just as for $\Delta E$ and the isotropic TDGL result for $n=d=2$
\cite{HQ2,HQ5}. $N_{d}$ and $\Delta E$ share approximately the
same time dependence with growth law exponent $z\approx2$. The energy
per dislocation is almost a constant, which means the dislocations
control the dynamics of the system.

\begin{figure}
\begin{center} \includegraphics[scale=0.24]{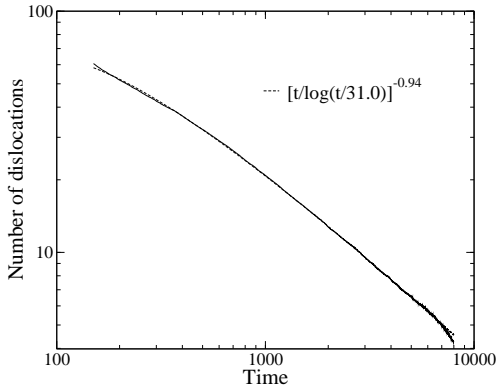}\end{center}

\caption{The number of dislocations $N_{d}$ for the anisotropic SH model
as a function of time averaged over 528 runs is given by the solid
line. There is an excellent fit to the form $(t/\log(t/a))^{-b}$
with $a=32.1$ and $b=0.94$. }
\end{figure}

\section{Analysis}

To quantify the extent of cross stripe migration of dislocations,
we measured the average distance $r_{\alpha}$ in the climb and glide
directions between two dislocations which are going to annihilate
with each other at time $t_{0}$. In Fig. 7, we show $r_{\alpha}$
v.s. $t-t_{0}$, where $t$ is the time we measure the distance. We
measured the components of average distance across and along the stripes.
In Fig. 7, we can see that at any $t-t_{0}$ the separation of two
annihilating dislocations across the stripes is much larger than the
distance along the stripes. This means the two dislocations tend to
approach each other along the direction across the stripes. We notice
that the average separation takes a power law form for the glide motion.
The climbing motion as one approaches annihilation is more complicated.

\begin{figure}
\begin{center}\includegraphics[scale=0.25]{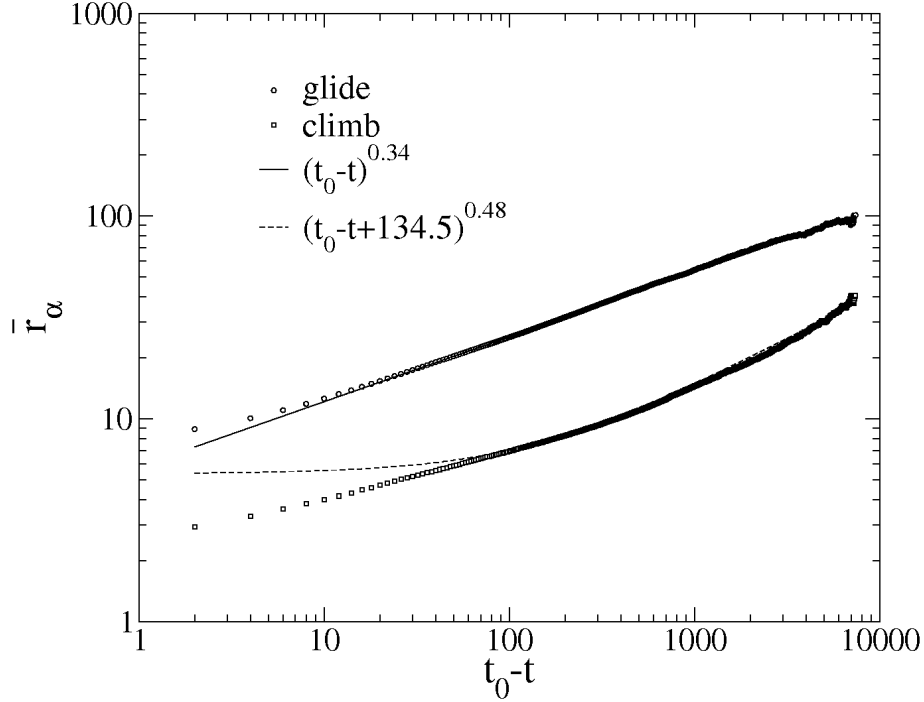}\end{center}

\caption{The average distance between two dislocations which are going to
annihilate with each other versus time before the annihilation. The
components across (glide) and along (climb) the stripes are measured
respectively.}
\end{figure}

\begin{figure}
\begin{center}\includegraphics[scale=0.31]{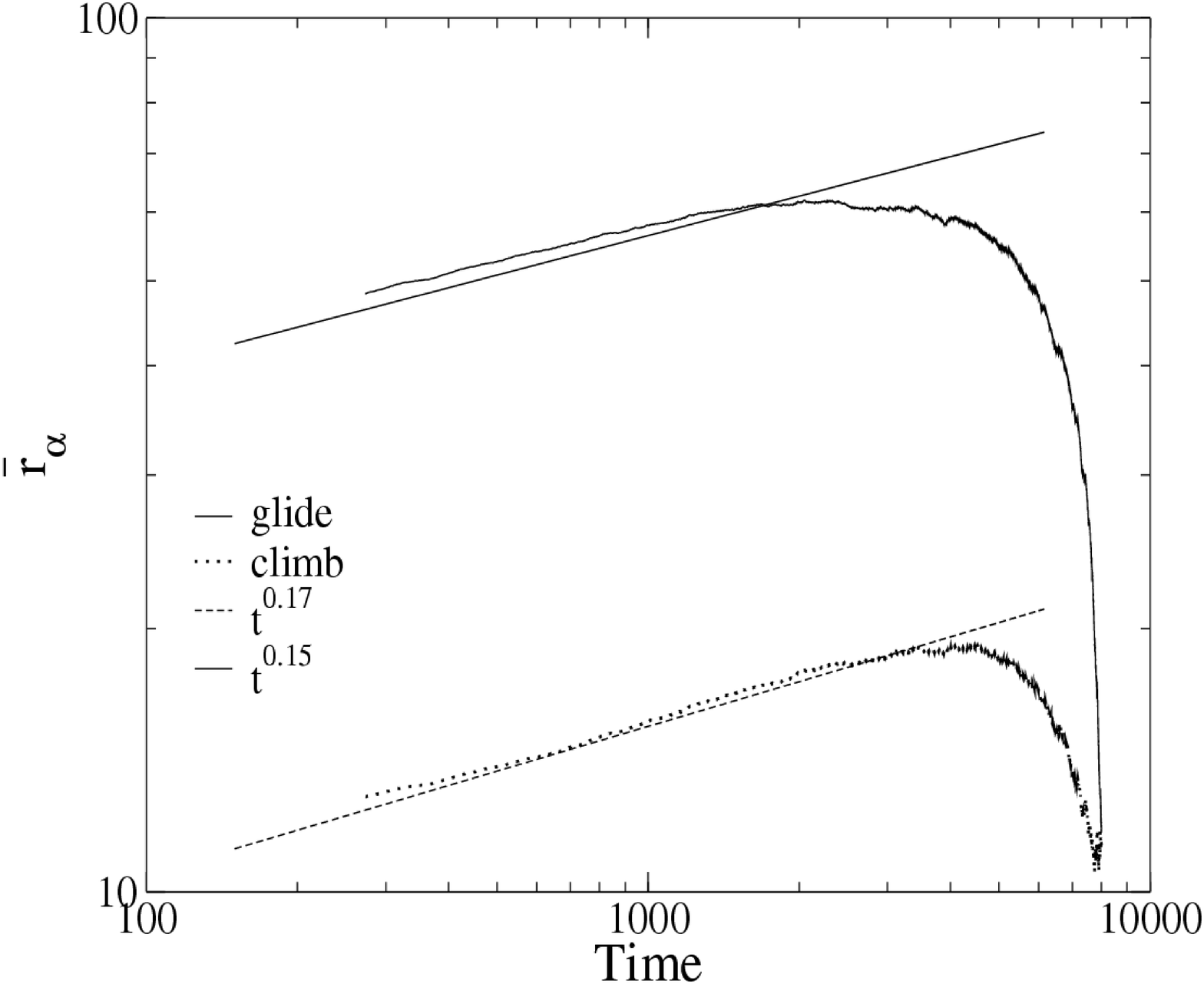}\end{center}

\caption{The average distance between two dislocations which are going to
annihilate versus the time after the quench. The components across
(glide) and along (climb) the stripes are measured respectively.}
\end{figure}

If we plot the average separation of defects heading toward annihilations
versus time after quench we obtain the result shown in Fig. 8. Unlike
in the NCOP TDGL case $\bar{r}$ deviates from a power-law behavior
for long times and does not serve as a good measure of the growth
law for the system.

In Ref.\cite{HQ5} we found good scaling results and a reasonable
theoretical model for describing the numerical results because there
was a simple scaling relation between the average separation $r$
of two annihilating defects and their relative speed $u$, $u\sim r^{-b}$.
The situation is more complicated here. In Fig. 9 we plot the average
relative speed in a given direction versus separation in that direction.
As the dislocations approach each other, their speeds increase. And
when the distance between them is too small, our measurement is unable
to follow the motions of the dislocations. So the data points in Fig.
9 with distances smaller than $r=16$ are not reliable and should
not be taken into account in the following analysis.

Using only the reliable data in Fig. 9 we obtain 
\begin{equation}
u_{\alpha}=A_{\alpha}r_{\alpha}^{-b}
\end{equation}
 where $b=0.75$ for both climb and glide, and $A_{g}=0.58$, $A_{c}=0.06$.
The assumption $\bar{u}_{\alpha}=\bar{u}_{\alpha}(r_{\alpha})$ is
only approximately correct. In fact $\bar{u}_{\alpha}$ depends on
both $x$-separation and $y$-separation distances, as is shown in
Fig. 10a. Correspondingly, there is a strong correlation as shown
in Fig. 10b between the glide separation $d_{g}$ and the separation
distance $d$. The climb distance is correlated with $d$ for small
enough separations.

\begin{figure}
\begin{center}\includegraphics[scale=0.39]{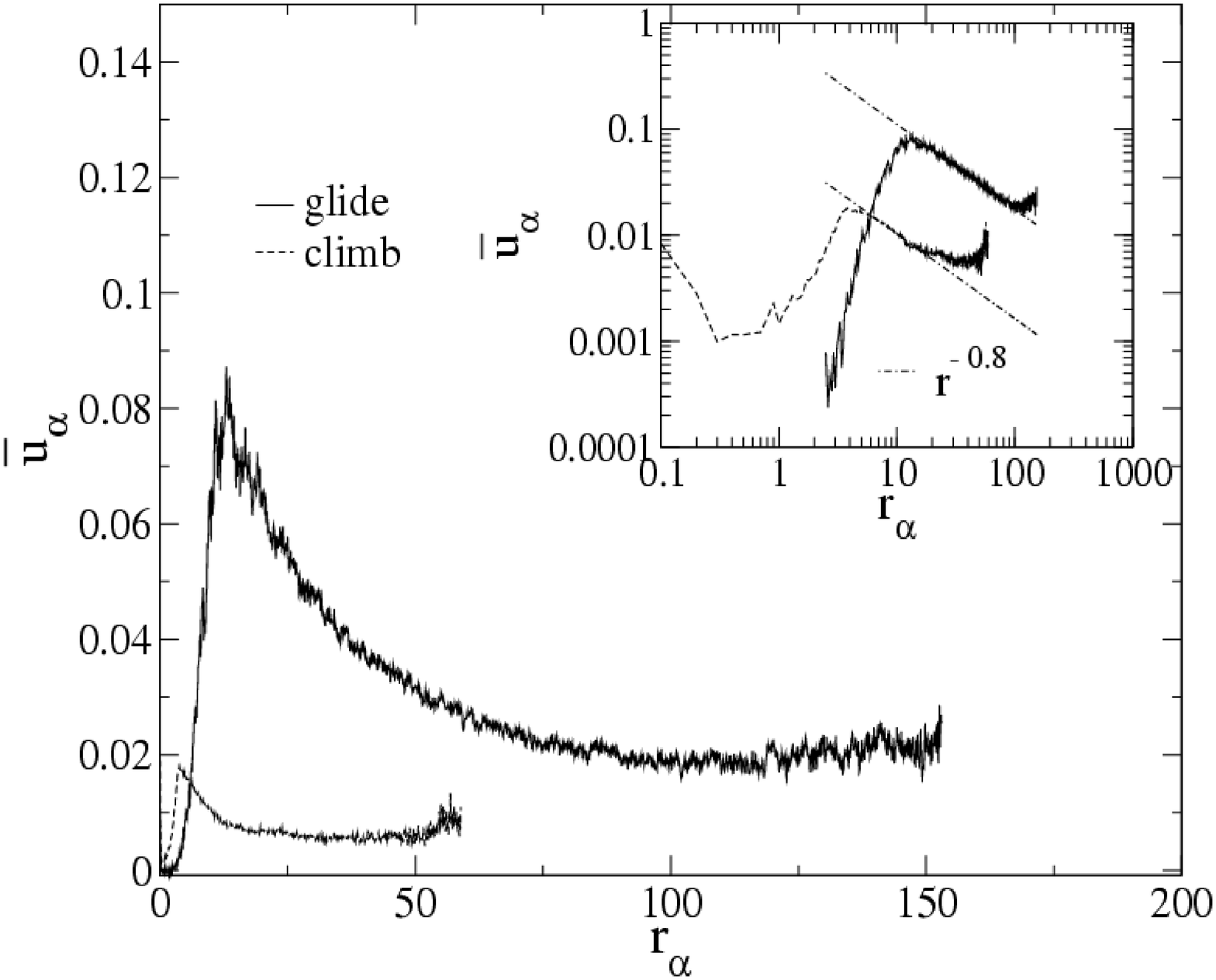}\end{center}

\caption{The average relative speed between two dislocations which are going
to annihilate versus the separation between them in that direction.
The best fit for each set of the glide data is given by $u(r_{\alpha})=A_{\alpha}r_{\alpha}^{-b}$
with $b_{\alpha}=0.75$ approximately for both glide and climb.}
\end{figure}

\begin{figure}
\begin{center} \includegraphics[scale=0.39]{eps/f10_1.eps} \includegraphics[scale=0.39]{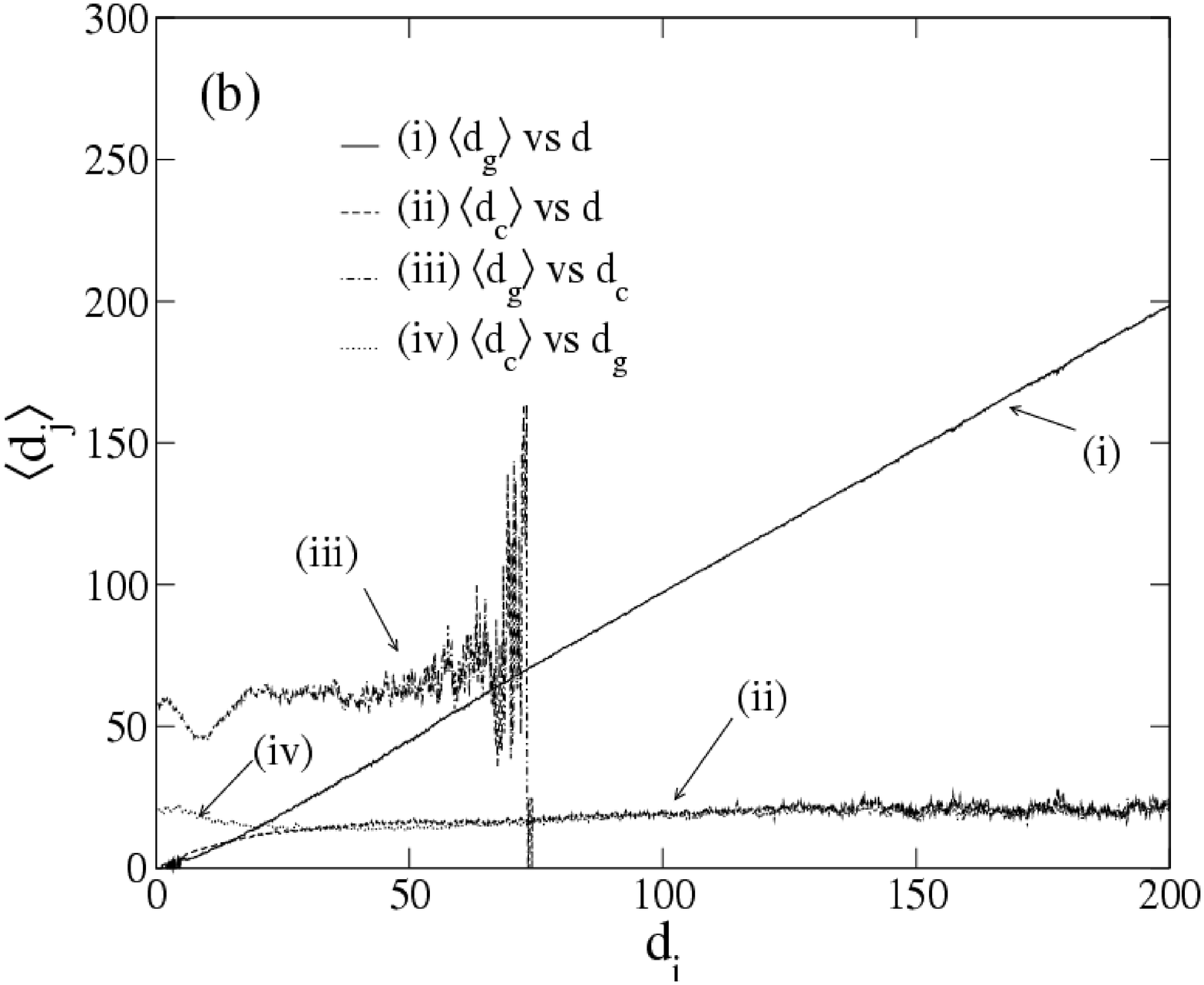}\end{center}

\caption{(a) The speed $u_{x}$ and $u_{y}$ depend on both $x$ and $y$.
(b) $d$ is the distance between two annihilating dislocations. $d_{g}$
is the separation distance across the stripes (glide motion). $d_{c}$
is the separation distance along the stripes (climb motion). And $d^{2}=d_{c}^{2}+d_{g}^{2}$.}
\end{figure}

If we plot the average relative speed versus time to annihilation,
Fig. 11, we obtain the approximate power-law results \begin{equation}
\bar{u}_{\alpha}(t)\sim(t_{0}-t)^{-1/z}\,\,,\end{equation}
 which is consistent with $z=2$ in both directions.

\begin{figure}
\begin{center}\includegraphics[scale=0.39]{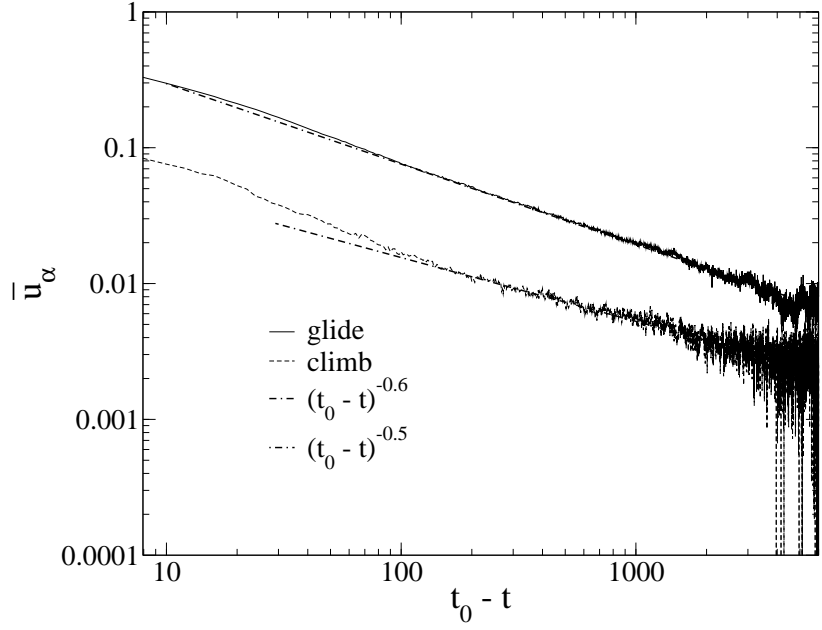}\end{center}

\caption{The average relative speed between two dislocations which are going
to annihilate versus the time before the annihilation in the glide
and climb directions.}
\end{figure}

\begin{figure}
\begin{center}\includegraphics[scale=0.39]{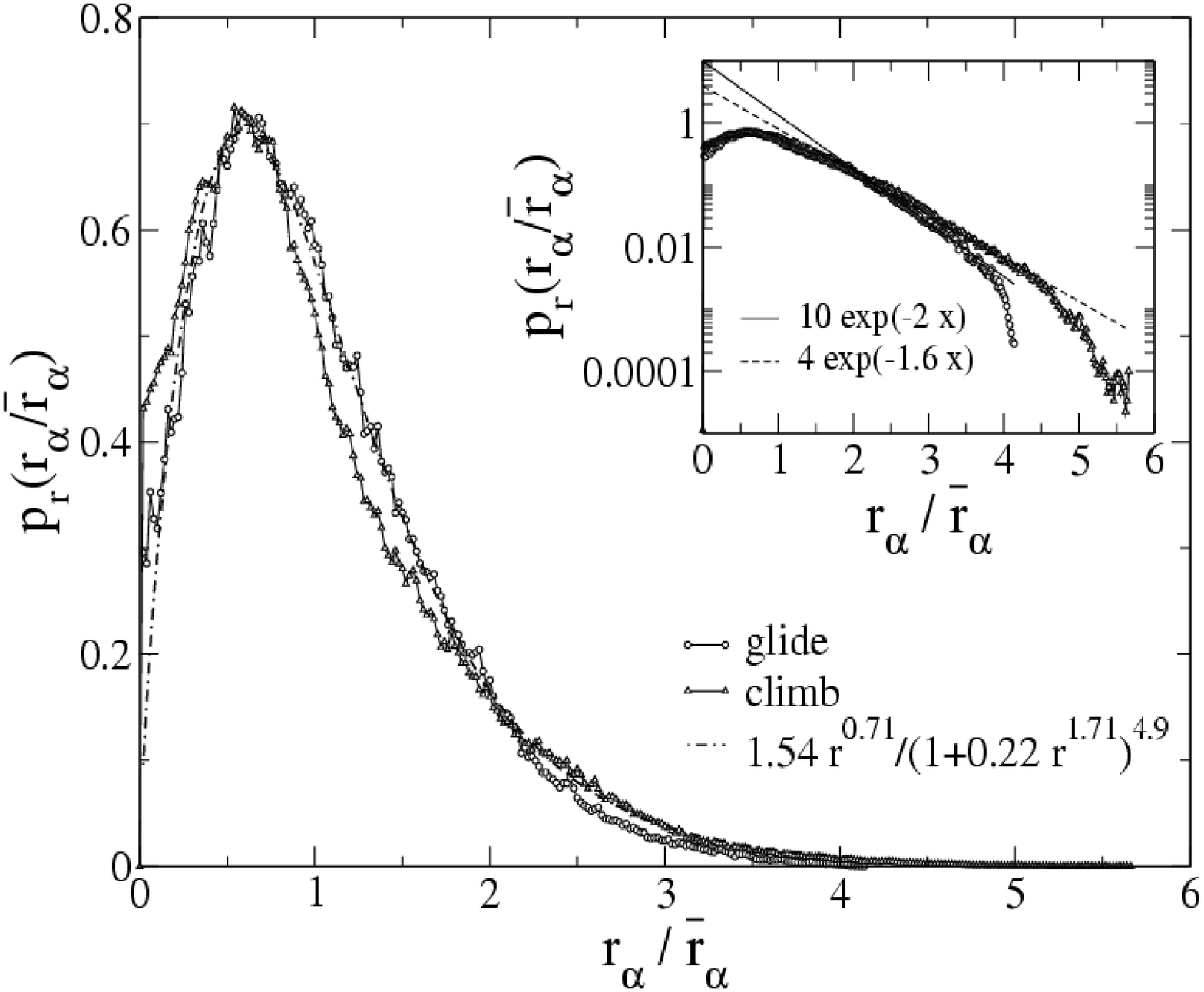}\end{center}

\caption{The separation probability distribution. The tail of the distribution
has an exponential form, as is shown in the insert. We fit the separation
distribution function with $y=a_{0}x^{b}/(1+a_{1}x^{1+b})^{c}$, where
$a_{0}$, $a_{1}$, $b$ and $c$ are parameters whose values are
given in the figure.}
\end{figure}

\section{Probability Distributions}

We can next turn to the associated probability distributions. The
first is the separation probability distribution. This is the probability
$P_{r}^{\alpha}(t)$ that at time $t$ two annihilating dislocations
are separated by a distance $r_{\alpha}$, where $\alpha$ corresponds
to climb or glide. We assume that $P_{r}^{\alpha}(t)$ takes a scaling
form if we plot it versus $r_{\alpha}/\bar{r}_{\alpha}(t)$ as shown
in Fig. 12, where the average $\bar{r}_{\alpha}$ is shown in Fig.
8. We obtain a reasonable scaling form but unlike in Ref. \cite{HQ5}
we do not find an algebraic large separation tail. Instead the scaling
function appears to decay exponentially. Notice that the scaling forms
are roughly independent of direction. We fit the separation distribution
function with $y=a_{0}x^{b}/(1+a_{1}x^{1+b})^{c}$, where $a_{0}$,
$a_{1}$, $b$ and $c$ are parameters. This function is the most
general form for the separation distribution obtained from \cite{HQ5}.
Since the fitting function has power-law decay tail, which is different
from the exponential decay of the real data, we are unable to fit
the large separation tail. When we fit the data, we restrict the value
of $c$ to be between $1$ and $5$. The resulting value of $c$ is
$4.9$. If we extend the upper limit on $c$, the best fit for $c$
increases. However we found that $b$ is always close to $0.7$ whatever
the upper limit of $c$.

We turn next to the statistics governing the relative speeds. The
average speeds are shown in Fig. 13a to be given approximately by
$t^{-1/2}$ for both directions in agreement with the scaling ideas.
If we fit the climb trajectories in a way that averages over the oscillations
in the climb component we obtain the average climb speed which is
in better agreement with the $t^{-1/2}$ result, as is shown in Fig.
13b.

\begin{figure}
\begin{center} \includegraphics[scale=.36]{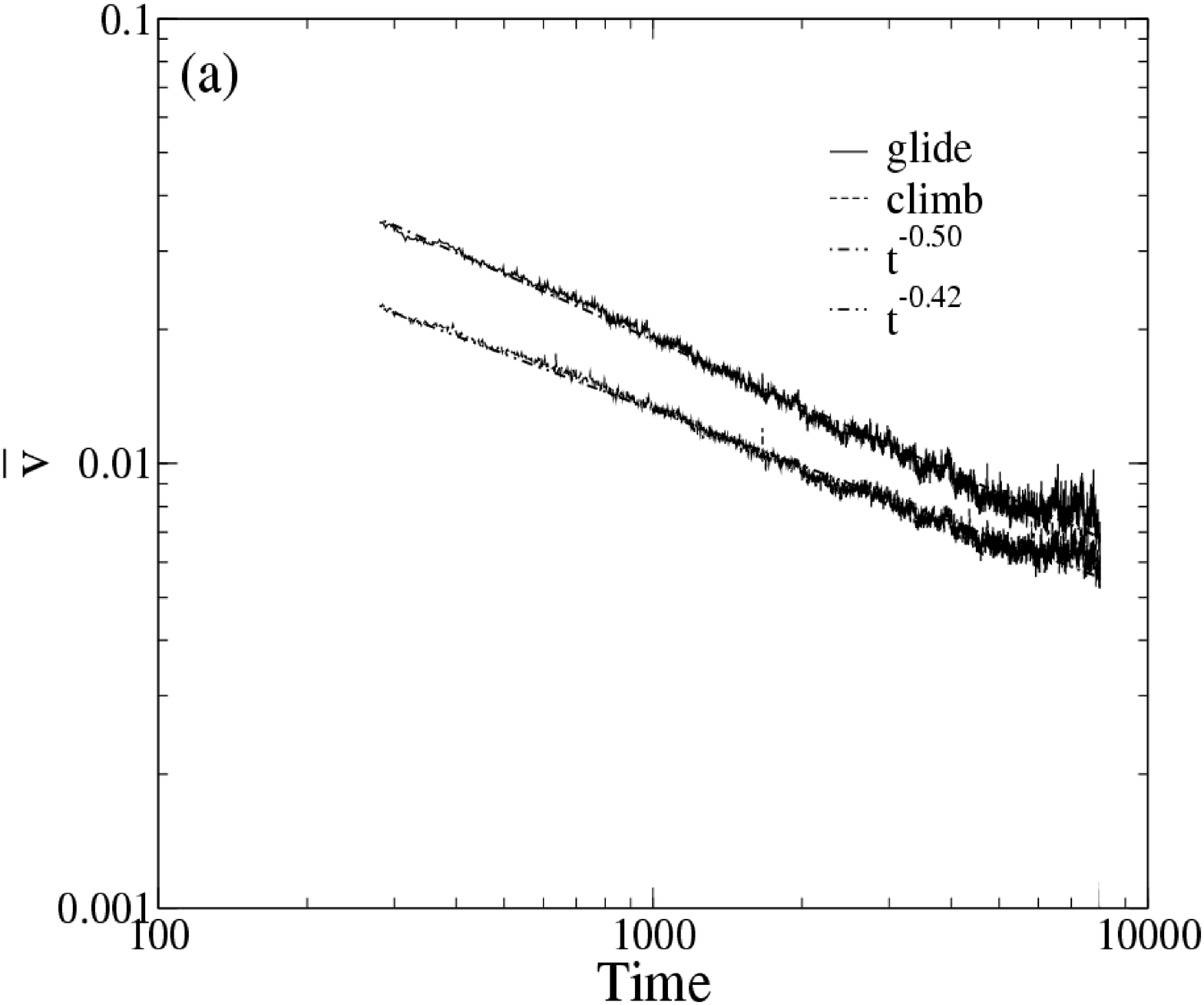} \includegraphics[scale=.31]{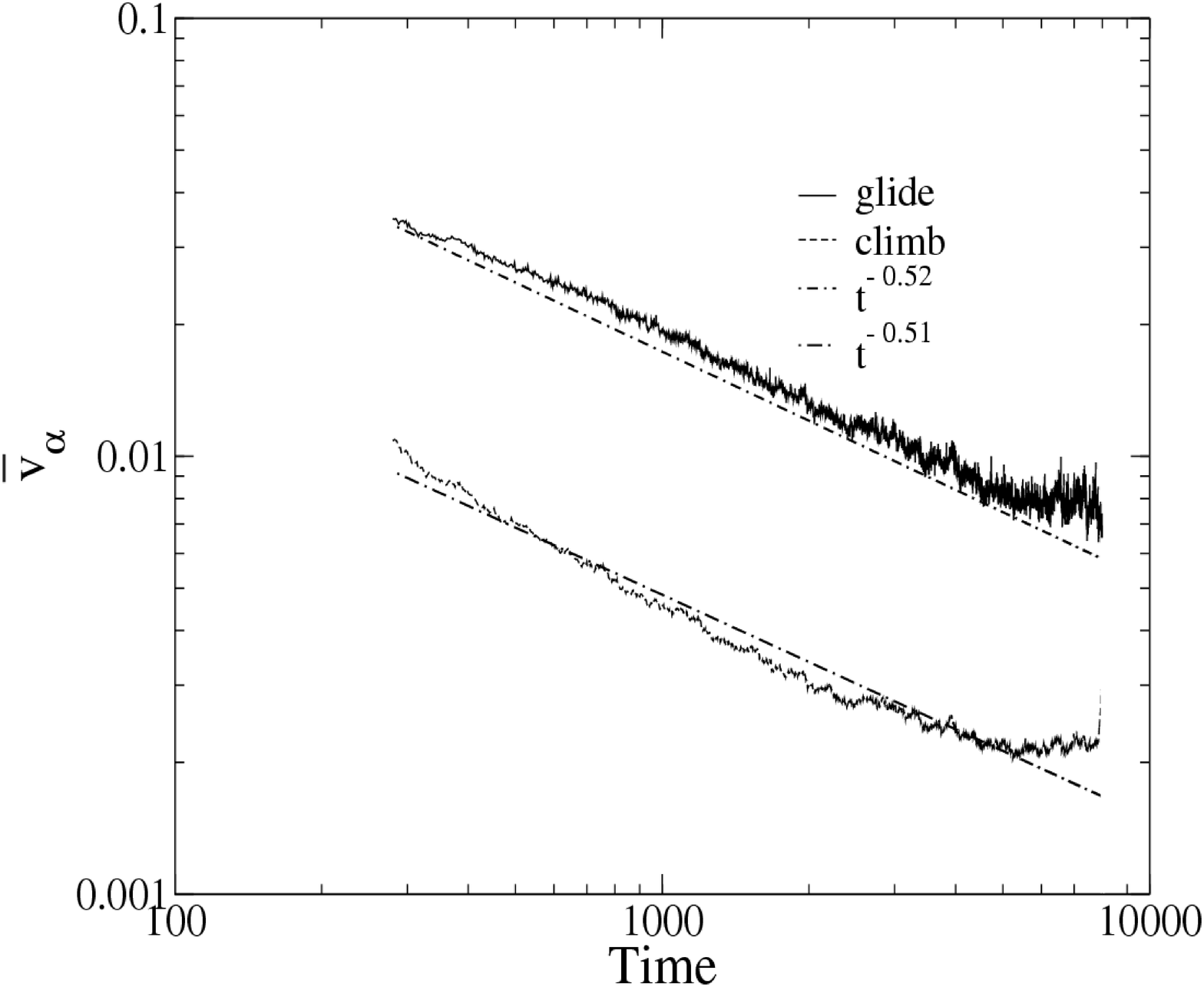}\end{center}

\caption{(a) The averages of the speeds, the transverse component (across
the stripes) and the longitudinal component (along the stripes). They
all obey simple power laws $\bar{v}_{\alpha}\sim t^{x_{\alpha}}$.
In (b), we use 10th degree polynomial to fit the trajectories of climb
motion and average out the oscillations on climb motion. The obtained
power-law exponent is closer to 0.5 in this case.}
\end{figure}

Finally we plot the speed distribution function in Fig. 14a. Clearly
we have large speed power-law tail. We find roughly that both glide
and climb motions have a $v^{-3}$ large speed power-law tail as shown
in Fig. 14b after we average out the oscillations on the $y$-component
(along the stripes). The distribution functions are sensitive to how
we treat the oscillations in the climb data. This may account for
the higher tail exponent 3.9 shown in Fig. 14a for the climb data.
We also use a quite general form for the speed distribution function
obtained from Ref.\cite{HQ5} to fit the data. The function is $y(x)=a_{0}x^{-2+(c(b+1)-1)/b}/(1+a_{1}x^{(b+1)/b})^{c}$.
We require that $y(0)$ to be nonzero. So we must have $-2+(c(b+1)-1)/b=0$.
The parameter $b$ is the same $b$ in Fig. 9, with a value 0.8. So
we have $c=(1+2b)/(1+b)=1.44$. So we fix the values of $c=1.44$
and $b=0.8$ in the curve fitting.

Again the scaling forms in the two directions are, to within our accuracy,
equivalent.

\begin{figure}
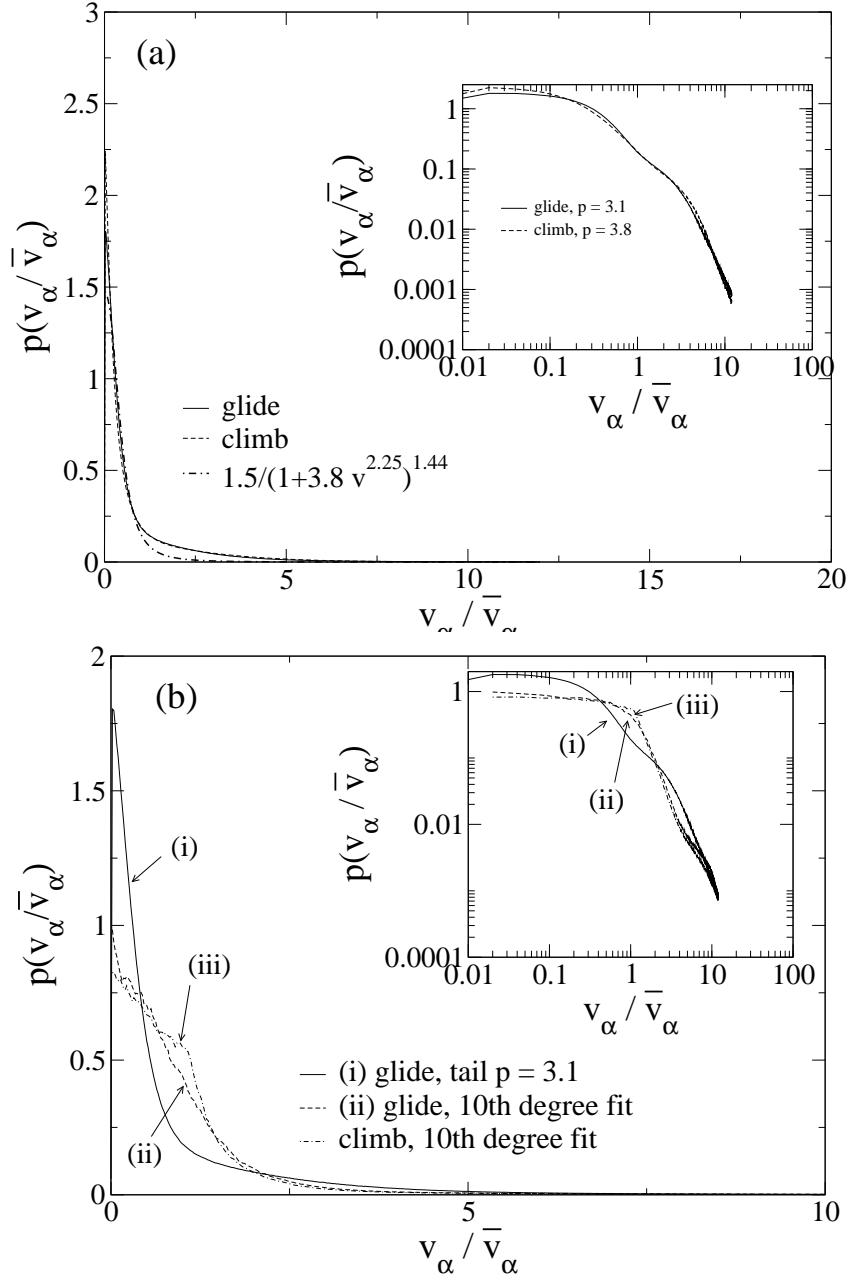

\begin{center} \includegraphics[scale=.4]{eps/f14_1.eps} \includegraphics[scale=.4]{eps/f14_2.eps}\end{center}

\caption{The distribution of the defects' speed for glide and climb directions.
After scaling the data at different times (from 300 to 8000), we collapse
all the speed data to one curve. The motions of gliding and climbing
are measured separately. In (b), the oscillations of the climbing
motion along the stripe are averaged out. We also use the 10th degree
polynomial fit to smooth the glide motion. }
\end{figure}

\section{Conclusion}

The kinetics of the anisotropic SH model are conceptually simpler
than for the isotropic SH case since there is only one disordering
defect. In the simplest picture one has a set of point defects ordering
in a fashion similar to a collection of vortices in an XY model \cite{HQ2}
but with anisotropic scaling. One has a scaling length $L_{\alpha}(t)\sim A_{\alpha}t^{x}$
and power-law behaviors for the decreasing energy, number of defects,
and average defect velocity with $x\sim0.5$.

The picture of annihilating point defects with a growth law of $t^{1/2}$
is roughly true for our anisotropic system and in this sense the system
is similar to the 2D XY model with the annihilation of vortices. Both
have bulk properties $\Delta E$ and $N_{d}$ which have the same
ordering time dependence. Both systems have large velocity power-law
tails which show $v^{-3}$ behavior. One simple result is that the
scaling functions $P(r)$ and $P(v)$ are nearly isotropic as shown
in Fig. 12 and 14. This is true despite the fact that the motion is
highly anisotropic. It appears that the annihilating defects organize
themselves in such a way that their relative velocity is radial. Once
this is true it is not important what the angle is between the relative
velocity and the stripes. As one looks closer the analogy breaks down.
The separation probability distribution shows a clear power-law tail
in the XY model but not in the anisotropic model. The average distance
between annihilation defects in the XY model serves as a good measure
of the growth law $\bar{r}\sim t^{1/2}$. This is not true in the
anisotropic model where $\bar{r}\sim t^{0.17}$. One interpretation
of our results is that the hypothesis of independent pairs of dislocations
breaks down at a much greater distance than for the XY model. It is
also possible that the independent pair mechanism used in Ref.\cite{HQ5}
works less well in this system. There may be correlations among different
pairs of dislocations. In any event our results can not be simply
explained by a rescaling of the glide and climb directions.

\vspace{3mm}

\textbf{Acknowledgements}: This work is supported by the Material
Science and Engineering Center through grant No. NSF DMR-9808595.


\begin{thebibliography}{11}
\bibitem{Harrison}C. Harrison, D. H. Adamson, Z. Cheng, J. M. Sebastian, S. Sethuraman,
D. A. Huse, R. A. Register and P. M. Chaikin, Science \textbf{290},
1558 (2000). 
\bibitem{Harrison2}C. Harrison, Z. Cheng, S. Sethuraman, D. A. Huse, P. M. Chaikin, D.
A. Vega, J. M. Sebastian, R. A. Register and D. H. Adamson, Phys.
Rev. E \textbf{66} 011706 (2002). 
\bibitem{CB98}J. J. Christensen and A. J. Bray, Phys. Rev. E \textbf{58}, 5364 (1998). 
\bibitem{HQ1}H. Qian And G. F. Mazenko, Phys. Rev. E \textbf{67}, 036102 (2003),
and the references therein. 
\bibitem{PismenASH}L.M. Pismen, Vortices in Nonlinear Fields, Oxford, London (1999),
page 63. 
\bibitem{Cross}G. Tesauro and M. C. Cross, Phys. Rev. A \textbf{34}, 1363 (1986). 
\bibitem{Goren}G. Goren, I Procaccia, S. Rasenat and V. Steinberg, Phys. Rev. Lett.
\textbf{63}, 1237 (1989). 
\bibitem{rasenat}S. Rasenat, V. Steinberg and I. Rehberg, Phys. Rev. A \textbf{42},
5998 (1990).
\bibitem{kramer}L. Kramer, E. Bodenschatz and W. Pesch, Phys. Rev. Lett. \textbf{64},
2588 (1990).
\bibitem{Braun}E. Braun and V. Steinberg, Europhys. Lett. \textbf{15}, 167 (1991). 
\bibitem{struc}E. Bodenschatz, W. Pesch and L. Kramer, Physica D \textbf{32}, 135
(1988). 
\bibitem{BP}E. Bodenschatz, W. Pesch and L. Kramer, Physica D \textbf{32} 5 (1988). 
\bibitem{boyer}D. Boyer, Phys. Rev. E \textbf{69}, 066111 (2004).
\bibitem{pesch}W. Pesch and L. Kramer, Z. Phys. B: Condens. Matter \textbf{63}, 121
(1986).
\bibitem{kamaga}C. Kamaga, F. Ibrahim and M. Dennin, Phys. Rev. E \textbf{69}, 066213
(2004).
\bibitem{Ward}W. Lopes private communication.
\bibitem{HQ5}H. Qian and G. F. Mazenko, Phys. Rev. E \textbf{70}, 031104 (2004).
\bibitem{HQ2}H. Qian and G. F. Mazenko, Phys. Rev. E \textbf{68}, 021109 (2003).
\bibitem{SH77}J. Swift and P. C. Hohenberg, Phys. Rev. A \textbf{15}, 319 (1977).
\end{thebibliography}
\end{document}